\documentclass[
    ,final            
  ]
  {aipproc}

\layoutstyle{6x9}

\begin{document}

\title{Metropolis Methods for Quantum Monte Carlo Simulations}

\author{D. M. Ceperley}{
  address={NCSA and Dept. of Physics\\
   University of Illinois Urbana-Champaign\\
    1110 W. Green St., Urbana, IL, 61801}
}

\begin{abstract}
Since its first description fifty years ago, the Metropolis Monte
Carlo method has been used in a variety of different ways for the
simulation of continuum quantum many-body systems. This paper will
consider some of the generalizations of the Metropolis algorithm
employed in quantum Monte Carlo: Variational Monte Carlo,
dynamical methods for projector monte carlo ({\it i.e.} diffusion
Monte Carlo with rejection), multilevel sampling in path integral
Monte Carlo, the sampling of permutations, cluster methods for
lattice models, the penalty method for coupled electron-ionic
systems and the Bayesian analysis of imaginary time correlation
functions.
\end{abstract}

\maketitle

\section{Introduction}

Though the original applications of the Metropolis {\it et al.}
method\cite{metropolis53} was to a classical system of hard disks,
the algorithm has since been found indispensable for many
different applications. In this talk, I will discuss some of these
applications of the Metropolis algorithm to quantum many-body
problems. This article will be strictly limited to the use of the
Metropolis rejection method within quantum Monte Carlo (QMC) and
not discuss other aspects of QMC. The richness of the Metropolis
algorithm, and the brevity of this article implies that I can only
touch briefly on a subset of these developments and must limit
myself to a superficial discussion. Others will discuss its use in
quantum lattice models, both of condensed matter and in lattice
gauge theory, so my focus will be on non-relativistic continuum
applications, in particular, on developments requiring
generalization of the basic Metropolis algorithm.  I will only
mention briefly the physics behind these applications, instead
referring to review articles.

We define the Metropolis algorithm as follows. Suppose $s$ is a
point in a phase space and we wish to sample the distribution
function $\pi (s)$. In the simplest algorithm, there is a single
transition probability: $T(s \rightarrow s')$. Later we will
generalize this to a menu of transition probabilities. One
proposes a move with probability $T( s \rightarrow s')$ and then
accepts or rejects the move, with an acceptance probability $A(s
\rightarrow s')$. Detailed balance and ergodicity are sufficient
to ensure that the random walk, after enough iterations, will
converge to $\pi (s)$, where by detailed balance we mean that:
\begin{equation}
\pi(s) T(s \rightarrow s') A(s \rightarrow s') =
 \pi(s') T(s' \rightarrow s) A(s' \rightarrow s). \label{DB}
\end{equation}
By ergodicity, we mean that there is a non-zero probability of
making a move from any state to any other state in a finite number
of moves. We refer to this class of algorithms as Markov Chain
Monte Carlo (MCMC) or Metropolis. The generalization of Metropolis
to non-uniform transitions was suggested by
Hastings\cite{hastings70}. Key defining features of MCMC are the
use of detailed balance to drive the distribution to a desired
equilibrium state and, in particular, the use of rejections to
achieve this.

In the ``classic'' Metropolis algorithm\cite{metropolis53}, the
state space is the 3N vector of coordinates of particles and the
distribution to be sampled is the classical Boltzmann
distribution, $\pi \propto exp(-\beta V(s))$. The moves consist of
single particle displacements chosen uniformly in a hypercube
centered around the current position and the moves are accepted
with probability given by $\min[1, \exp(-\beta (V(s')-V(s) )]$. In
the hard sphere system, it comes down to determining whether there
are overlapping spheres in the proposed move.

How can this simple Metropolis algorithm be generalized? First of
all for quantum system not only does the distribution need to be
sampled, the many-body wavefunction is unknown. This is done by
augmenting the spatial coordinates with an imaginary time, thereby
mapping the $d$ dimensional quantum system onto a $d+1$
dimensional space, which is then sampled with MC. In some
situations the space needs further enlarging, for example to add a
permutation of particles in taking into account Bose or Fermi
statistics.  The biggest problem with the Metropolis method is
that the approach to equilibrium and the autocorrelation of
properties can be very slow. So another important generalization
is in finding new, more efficient ways of moving through the
sample space. In the case of the penalty method, discussed below,
we allow noisy and cheap estimates of the potential energy
function. Finally, the Bayesian methods use MCMC to extract the
last bit of information from data generated by another MCMC
simulation, {\it i. e. } for data analysis.

There are several themes of this article. First, it is fruitful to
generalize MCMC to much more complicated systems. Second, without
the ability to generalize, much of quantum many-body physics would
still not be accessible to theory. Finally, the algorithmic
techniques easily cross disciplinary boundaries, from condensed
matter, to high energy, nuclear and chemical physics and more
recently into statistics and economics, making a meeting such as
this important. The Metropolis technique is indeed perhaps the
most powerful computational method (the criterion being that there
are no alternative algorithms) and has become the computational
common denominator amongst those dealing with many-body systems.
Of course, this generality implies that there have been many
developments of which I can only touch on a few. Others will be
covered elsewhere in this conference.

\section{Variational Monte Carlo}

The first major application of Quantum Monte Carlo (QMC) to a
many-body system was by McMillan\cite{mcmillan65} to liquid
$^4$He. Previous calculations were for only a few particles and
are discussed by Kalos elsewhere in this volume. The method
McMillan used is now referred to as Variational Monte Carlo and is
based on the familiar variational method for solving ground state
quantum problems. As often happens, there was a simultaneous
calculation by Levesque et al.\cite{lev65} never published.

Assuming an appropriate trial function $\psi_T(R;a)$ where $R$ are
the particle coordinates possibly including spin, and $a$ is a set
of parameters, the variational energy,
\begin{equation}
E_V = \frac{\int dR \psi_T^* {\hat H} \psi_T}{\int dR |\psi_T|^2}
= \langle \psi_T^{-1} {\hat H} \psi_T \rangle
\end{equation}
is an upper bound to the ground state energy. In the second
equality, $\langle \cdots \rangle$ denotes an average over the
probability distribution of the trial function $|\psi_T|^2$. It is
in generating this distribution, that Metropolis Monte Carlo comes
into play. Traditionally, in the variational method the above
integrals are done deterministically, severely limiting the form
of the trial function. However, using MCMC, any distribution can
be sampled including those having explicit particle correlations.
In fact, it is an consequence of the large value of the
electron-nuclear mass ratio that mean field methods are so
pervasive in solid state physics and quantum chemistry.  Such
methods are much less efficacious for liquid $^4$He and other
important correlated quantum systems.

McMillan in his pioneering work on the simulation of liquid $^4$He
used the pair product (Jastrow) trial function:
\begin{equation}
\psi_2(R)= \exp[-\sum_{i<j} u(r_{ij};a )].
\end{equation}
This form is identical to that of a classical Boltzmann
distribution for a system of atoms interacting with a pair
potential, such as a rare gas liquid if we make the substitution
$u(r) \rightarrow v(r)/k_BT$. His work and many that have
followed, have established that this wavefunction provides a
physically correct, though not exact, description of the ground
state of liquid helium. Though the code that computed the
distribution was thereby identical to a classical code, McMillan
had to introduce new features in the algorithm in order to
calculate properties such as the kinetic energy and momentum
distribution and to optimize parameters in the trial function.

To treat fermion systems, such as liquid $^3$He, one generalizes
the trial function by multiplying by a Slater determinant of one
body orbitals, a Slater-Jastrow trial function. There was a twelve
year lag in generalizing VMC to fermion systems, primarily because
of a psychological barrier. The perception in the community was
that fermion MC would be too slow and the algorithm could be
non-ergodic if the determinant were to be included in the
sampling, so instead, approximate methods were introduced take
into account antisymmetry.

Concerning the question of efficiency, it takes order $N^3$
operations to evaluate a determinant. However, it turns out there
is an algorithm, the Sherman-Morisson formula, discovered when
determinants were evaluated by hand, that allows one to change a
single column or row in only $N^2$ operations\cite{C003}. Using
the update method with Slater determinants, it is possible to do
single particle moves within MCMC quite efficiently.  In fact,
until one reaches several hundred fermions, the time per Monte
Carlo step is not much slower than MCMC for the equivalent bosonic
system. One implication for the algorithm is that rejections are
much ``cheaper'' then acceptance. As a result efficient VMC
fermion calculations have larger step sizes and much smaller
average acceptance ratios than usual.

The other psychological barrier concerned the non-classical
distribution when a squared determinant is present in the trial
function. If the trial function is real, or describes a closed
shell (non-degenerate ground state), the phase space will be
divided into regions separated by the nodes of the trial function.
It seems possible that the random walk could get trapped in one
pocket of phase space, giving a bias to computed properties.
However, once a Metropolis random walk with a determinant was
attempted, it was observed that the sign of the trial function
changed every few steps. The issue of ergodicity was finally
settled with the proof of the tiling theorem\cite{C059}: for
determinants coming from the solution of mean field theory, all
the various pockets are equivalent, so that even if the random
walk remained in one pocket, the results would be unbiased.

A side effect of attempting to simulate fermion systems was a
generalization of MCMC, known as ``force-bias'' MC. In MCMC there
is always a struggle between moves with high acceptance rates that
go long distances and moves that are fast\footnote{This
illustrates richness of the Metropolis method; one has great
freedom in attempting moves.  In molecular dynamics, once you
specify the Lagrangian you are more or less stuck with the
resulting dynamics.}. Until the update formula allowing much
faster single particle moves was employed,  $N$-particle moves
were used. However, as $N$ becomes large, one is forced to take
increasingly smaller step sizes to get reasonable acceptance
rates. Note that in better wavefunctions such as backflow, all
rows depend on all of the particle coordinates, so update methods
are not useful and non-uniform methods are needed for these more
accurate VMC calculations. Kalos suggested\cite{C003} that a
non-uniform transition probability, $T( s \rightarrow s')$ which
locally approximates the equilibrium distribution will have a
larger acceptance rate, and hence will allow a larger step size
and faster convergence. (Note that this followed the work of
Hastings\cite{hastings70}.) These ``directed'' improvements to the
classic Metropolis were soon picked up for simulations of
classical liquids giving rise to what was called ``force-biased''
Monte Carlo\cite{pangali78} and ``smart'' Monte
Carlo\cite{rossky78}.

In the ``smart'' Monte Carlo algorithm, a form that will also
appear in diffusion MC, one uses an offset Gaussian:
 \begin{equation}
 T(s \rightarrow s') = C exp( -(s' - s - D \nabla ln (\pi) )^2/ 4D
 )
 \end{equation}
where $\surd D$ is the step size and $C$ a normalization constant.
Then the new move is accepted with probability equal to:
\begin{equation}
A(s \rightarrow s') = min[ 1, \frac{ T(s' \rightarrow s)\Pi(s')} {
T(s \rightarrow s')\Pi(s)} ]
\end{equation}
By making a Taylor expansion about the current positions, one can
verify that the acceptance probability deviates from unity only by
terms second order in $s'-s$.

Another generalization of the Metropolis algorithm within VMC
concerns the properties of a bosonic solid\cite{C004}. An accurate
trial function for the quantum solid is a symmetrized product of
localized functions:
\begin{equation}
  \psi_T = \Psi_2(R) \frac{1}{\sqrt{N!}}\sum_P \prod_i \phi (r_i-Z_{P_i})
\end{equation}
where $P$ is a permutation and $\phi(r)$ is a localized orbital,
{\it e. g.} $\exp(-c r^2)$ , and $Z_i$ is the set of lattice
sites. Mathematically the solid wavefunction is a permanent, which
is very slow to evaluate explicitly once N gets large with an
operation count proportional to $N~2^N$ . However, it is easy to
add the permutation to the sample space since the trial
wavefunction is non-negative for all permutations. To sample the
combined space, we need to add a transition move to change the
permutation. Pair permutations are sufficient to move through the
entire space of $N!$ permutations. This way of generalizing
Metropolis had a spin-off into another area of quantum physics:
analyzing the debris produced by colliding bosons in particle
accelerators where, again, one must symmetrize over assignment of
particle labels\cite{zajc}. The idea of using Metropolis
simultaneously in both a continuous and discrete space turned out
to be essential in the VMC simulation of nuclei. Lomnitz-Alder et
al. \cite{vj81} sampled the ordering of spin and tensor operators
in a VMC calculation of small nuclei by executing a random walk in
the operator ordering along with normal moves of the positions of
the nucleons.

For inhomogeneous systems, the above form of trial function of a
crystal is awkward since it requires prior knowledge of the
crystal sites. In the {\it shadow wavefunction}\cite{vitiello} for
each real particle, one adds a complimentary shadow particle, also
with a Jastrow correlation and coupled to the real particles. The
shadow particle acts like the lattice site, but can move to find
its optimal position. This function can be considered a cousin to
path integrals, discussed below. The shadow trial function
spontaneously orders, so one does not have to specify the lattice
beforehand. It gives lower energy at the expense of additional
integrals, which don't cost much in VMC anyway. Without the MCMC
method, this wave function, which is quite accurate for a variety
of problems, would be unlikely to be considered.

One issue that arises for these more complicated trial functions
is how to move through the space efficiently. The simplest way to
implement the random walk is through a ``menu'' of moves, where
first one takes a particle coordinate move and then a
permutational change move (or a shadow move). One generalizes Eq.
1 to enforce detailed balance for each move separately. However,
the product of such probabilities, needed to have a Markov
process, does not, in general, satisfy detailed balance. But this
is not important; it is sufficient that the desired stationary
state be an eigenfunction of each of the possible menu items. The
freedom to depart from strict detailed balance is invaluable. This
was commonly known\cite{hastings70} in the early days though
rarely discussed.

\section{Brownian Dynamics and Diffusion Monte Carlo}

It is often stated that one does not use MC to study dynamical
problems. However, early on, MCMC was used to study kinetic
phenomena in the Ising model\cite{kalosbinder}.  Around 1977, we
began doing simulations of polymer dynamics\cite{C005,C016}. A
polymer in a solvent moves by Brownian motion. If hydrodynamic
forces are ignored then the master equation for the probability
density is the Smoluchowski equation:
\begin{equation}
 \frac{df(R,t)}{dt} = D\nabla[ \nabla+ \beta \nabla V(R) ]
  f(R;t).\label{BD}
\end{equation}
Here $f(R;t)$ is the distribution function for the system at time
$t$ and $D$ is the diffusion constant. Rossky et al.
\cite{rossky78} observed that the smart MC algorithm is equivalent
to one step of the Langevin equation assuming that the velocity is
randomized at each step. This randomization occurs physically
because a large blob of the polymer gets frequently hit by the
solvent during each time interval.

A problem we encountered in the simulation of the Brownian
dynamics was that it was necessary to take a very small ``time
step'' in order to get the proper equilibrium distribution,
because two particles would overlap in a region where the
interparticle potential was highly non-linear and the subsequent
step they would be thrown into a completely unphysical region.
Initially, we would solve this with a kludge; by putting an upper
limit on the force. However, after our experience with smart MC
with VMC, we decided to enforce detailed balance at each step, by
accepting or rejecting \`a la Metropolis. Detailed balance is an
property of the exact solution of Eq. \ref{BD}, and quite easy to
enforce with rejections. This allows better scaling in the number
of particles and a convenient way to decide on what time step to
take.  Of course, the equilibrium distribution will be exactly the
Boltzmann distribution for any time step, but to get realistic
dynamics, we adjusted the time step to get the average acceptance
ratio greater than 90\%. One can approximately correct for the
effect of rejections by rescaling the time in order to get the
exact diffusion constant. This use of rejection in Brownian
dynamics was a precursor to hybrid methods latter developed within
lattice gauge theory\cite{hybrid}. In hybrid methods, one
typically takes multiple dynamical steps before deciding whether
to accept the trajectory.

Now returning to quantum mechanics, it is highly desirable to go
beyond variational MC, since it is difficult to get more out of
the simulation than is put into the trial wavefunction. One needs
a more automatic scheme, where the stochastic process generates
the distribution. Such an approach, attributed to Fermi and
Wigner, who realized that the non-relativistic Schrodinger
equation in imaginary time is a random walk, and in the limit of
large imaginary time gives the ground state of the quantum system
\begin{equation}
-\frac{d \phi(R;t)}{dt}= [- D \nabla^2  +V(R) ]
\phi(R;t)\label{PMC}
\end{equation}
where $D= \hbar^2/2m$. The first many-body application of this
approach (GFMC) by Kalos {\it et al.}\cite{kal74} is not a
Metropolis method but a zero time step error method, in which
imaginary time is sampled. GFMC is based on the integral equation
formulation of the eigenvalue problem. Related methods have been
reintroduced in recent years in quantum lattice models under the
names of continuous time world-line algorithms\cite{prok96} as
discussed elsewhere in this volume by Troyer.

After working on the dynamics of polymers, the similarity between
the GFMC approach and that of Brownian Dynamics was apparent. The
connection is made by applying importance sampling to Eq.
\ref{PMC}: multiply the equation by a trial function $\Psi_T$ and
rewrite in terms of $f=\phi \Psi_T$. The resulting master equation
for $f$ is:
\begin{equation}
   -\frac{df(R;t)}{dt} = [-D \nabla^2+2 D \nabla (\nabla \ln
   |\psi_T|)
    + \psi_T^{-1} {\hat H } \psi_T ]f(r;t).
\end{equation}
This is the same as Eq. \ref{BD} for Brownian dynamics except for
the addition of the last term which is a branching process already
familiar within GFMC. The process of solving the Schrodinger
equation  with a drift, diffusion and branching process is known
as diffusion Monte Carlo (DMC). The values of imaginary times are
not sampled as in GFMC, so the method does have a time step error;
however the concepts of detailed balance and Metropolis rejection
are applicable since the exact evolution satisfies:
 \begin{equation}
 |\Psi_T (R)|^2 f(R \rightarrow R';t) = |\Psi_T (R')|^2 f(R' \rightarrow
 R;t).
 \end{equation}
This is the familiar detailed balance equation, however with a
subtle difference: $f(R \rightarrow R';t)$ is not a normalized
p.d.f. because of the branching process. Nonetheless, adding a
rejection step is both possible and highly desirable and results
in faster convergence at large $N$ as compared with GFMC
\footnote{There is a problem with the combination of branching and
rejection: that of persistent configurations. If there is a region
of phase space where $q e^{-\tau (E_L -E_T)} > 1$ then the
algorithm is unstable.}. More importantly it is simpler and easier
to integrate with fixed-node method \cite{and75} for treating
Fermi statistics and is now the almost universal choice for zero
temperature quantum problems.

Methods higher order in the time step have been occasionally
investigated to simulate the diffusion Monte Carlo
equation\cite{helfand}. However, if they rely on expansions of the
trial function or potential, the algorithm will fail badly at
non-analytic points, such as when two particles get close
together. As a consequence, higher order methods have not been
very successful for continuum electronic systems. The approach
based on rejections has a different principle than expansion in
the timestep, namely detailed balance is put in.  The equivalent
in deterministic algorithms such as molecular dynamics, is
reversibility in time. The crucial question is not the integration
order, but how much computer time it takes to get the error to a
certain accuracy. By enforcing detailed balance in DMC, one gets
the correct result both as the time step goes to zero and as the
trial wavefunction gets more accurate.

The DMC method had a spectacular debut\cite{C015}, in a paper
which has the highest citation count of any simulation. The
citations were not for introducing the DMC method, but because the
energy of a homogenous system of electrons is taken as the
reference system in density functional calculations of molecules
and solids. The details of the DMC method are given in ref.
\cite{C025}. The Metropolis algorithm using rejection was crucial
in giving accurate results for a variety of densities, phases and
particle numbers.  A recent review of applications obtained with
DMC are described in Foulkes {\it et al.}\cite{fou00}.


\section{Path integral methods}

The previous sections described applications at zero temperature.
We now consider the finite temperature Path Integral Monte Carlo
algorithms. In the same year as the celebrated Metropolis paper,
Feynman\cite{feynman53} showed that bosonic systems in equilibrium
are mathematically isomorphic to classical ``polymer-like''
systems. A single classical particle turns into a ring polymer
$r(t)$ where $t$ the time index is in the range $0 \leq t \leq
\beta=(k_B T)^{-1}$. In discrete-time path integrals, the time
index has only discrete values, $t_k = k \tau$ with $1 \leq k \leq
M$, $\tau$ is the time step, and $k$ the ``Trotter'' index. The
equivalent to the classical potential of the ``polymer'' is a sum
of the ``spring'' terms (from the quantum kinetic energy) and the
potential energy and called the action:
\begin{equation}
S(R(T))=\sum_{i=1}^M \frac{(r_i-r_{i+1})^2}{4 D \tau}  + \tau
V(r_i).
\end{equation}
An important aspect of equilibrium paths is that they are periodic
in imaginary time, which means for distinguishable particles $r
(t+\beta)=r(t)$.

The most interesting applications of PIMC involve systems with
Bose and Fermi statistics. There one must symmetrize over the
particle labels by allowing the paths to close on a permutation of
themselves. For Bose systems, all permutations have a positive
contribution. In Feynman's theory\cite{feynman53}, it was the
onset of a macroscopic permutation which was responsible for the
phase transition of liquid $^4$He at low temperature and the
observable properties of Bose condensation and superfluidity. The
MCMC simulations were crucial\cite{griffin} in finally convincing
most of the low temperature community that BEC really is
responsible for superfluidity and that Feynman had it right after
all. Applications to PIMC to helium are discussed at length in ref
\cite{C095}.

There was a twenty-five year lag between the development of
Metropolis MC and Feynman path integrals and the large scale
computer applications of PIMC.  There was considerable small-scale
work during this period, much of it unpublished (see
Jacucci\cite{jacucci84}), but not until the late 1970's did this
mature into computational efforts attacking important physical
systems, in the lattice gauge community\cite{creutz79}, in
chemical physics\cite{barker} and in lattice models for solid
state physics\cite{suzuki77a,scalapino81,deraedt81}. This 25 year
lag was most likely due to a lack of access to sufficiently
powerful facilities, combined with use of the basic Metropolis
algorithm, which is notoriously inefficient for path integrals.


There were  some problems to overcome in using MCMC to simulate
bosonic superfluids or indeed any system at a low
temperature\cite{C029,C035}. The primary problem is similar to
that encountered in polymer simulations; namely, that as a polymer
gets longer the correlation time of the random walk increases. One
can show\cite{C095} that the efficiency of any MC  with local
moves will drop at least as fast as $M^{-3}$ even for free
particles. This leads to a bad scaling versus the number of time
slices. There are several overlapping solutions to this problem.
At the technical level, one wants the best feasible approximation
for the action so as to avoid the necessity of many time slices
({\it i. e. } large $M$). Though this is an important topic, it is
irrelevant for the purposes of this lecture. The other approach is
to try to optimize how the state is changed: the transition
probability.

Essentially, the problem is how to move a chunk of the path
together. In order to get a permutation move accepted of $n$
atoms, in liquid $^4$He, it is essential to move a substantial
section, typically 8 time slices of the $n$ atoms involved. (Hence
one needs to move 24 coordinates at once.)  One wants to sample
gross features of the change and then sample finer details. The
idea is not to waste time computing details until gross features
are shown to be reasonable. There were several attempts to make an
efficient algorithm including the ``staging'' method\cite{staging}
and a method inspired by diffusion Monte Carlo\cite{C029}. Here we
will sketch the most successful generalization, multilevel
sampling\cite{C035,C095}.

In the multilevel method, a move is partitioned into $\ell$ levels
with an approximate ``action'' or distribution function for that
level $\pi_k (s)$ with the requirement that the level action equal
the true action at the highest level: $\pi_{\ell} (s) = \pi (s)$.
One samples the trial variables at each level according to some
probability distribution, $T(s_k')$.  Then those variables are
tentatively accepted with a generalized Metropolis formula,
ensuring detailed balance at each level:
 \begin{equation} A_k=
\min[ 1 , \frac{ T_k (s) \pi_k(s') \pi_{k-1}(s)}{ T_k (s')
\pi_k(s) \pi_{k-1}(s')} ]. \end{equation}
 If a move is rejected at
any level, one returns to the lowest level and constructs a
completely new move. This algorithm is applied to path integrals
by first, sampling the midpoint of the path, and with a certain
probability, continue the construction, sampling the midpoints of
the midpoints, etc. One gains in efficiency because the most
likely rejection will occur at the first level, when only
$2^{-\ell}$ of the computational work has been done.

Now consider the problem of how to carry out the bisection; in
other words, how to sample the midpoint of a Brownian bridge.
Given two fixed end points of the bridge, $R_0$ at time $0$ and
$R_{\beta}$ at time $\beta$, what is the distribution of a point
on the bridge $R_t$ at time $t$ with $0 < t < \beta$? For free
Brownian motion this was solved by L\'evy. For any quantum system,
the probability distribution of $R_t$ is:
 \begin{equation}
  T(R_t) = \frac{ \langle R_0 | e^{-t {\hat H} }| R_t \rangle
                   \langle R_t | e^{-(\beta-t) {\hat H}} | R_{\beta} \rangle }
                 { \langle R_0 | e^{-\beta {\hat H}} | R_{\beta} \rangle
                 }.
  \end{equation}
In this respect, path integrals are simpler than polymers, since
the action only contains terms local in imaginary time. For free
particles $T(R_t)$ is a Gaussian with an easily computed mean and
variance. For interacting particles, one needs to approximate this
by a Gaussian, with a mean and covariance perturbed from the free
particle values by the interaction with neighboring
particles\cite{C095}: an approach similar to smart Monte Carlo.

To start off the multilevel sampling, one first samples the
permutation change. In the case of the quantum crystal described
in the previous section, we were only interested in small local
permutations. However in a superfluid, one is particularly
interested in permutation changes which span the entire cell.  To
get winding number changes, which changes the superfluid density,
you need to construct a permutation which can cross the entire
system. That is, the cycle length needed to make a change in the
winding number is roughly equal to $N^{1/3}$ in 3D. Such a
permutation can be found with a random walk in index
space\cite{C095}. Once the permutation is established, then the
actual path is constructed.

The multi-level method has been independently discovered several
times. A recent example is the technique of pre-rejection used for
classical simulations\cite{gelb}. Suppose one computes a empirical
pair potential first and then a more complicated potential using
LDA after the first screening is done. If one has an accurate
empirical potential, one can quickly make many large displacement
moves, perhaps with a fairly low acceptance probability, and then
only on those rare moves that make it through the first level, go
to the expense of computing the accurate potential. Multilevel
sampling can also be used to improve the efficiency of
VMC\cite{dew00}. There are also similar ideas developed in the
polymer world as described in Frenkel's contribution.

For Fermi statistics, one subtracts the sum of the odd
permutations from that contribution of even permutations, leading
to the infamous fermion sign problem.  We are still struggling
with this problem today. One approach is to restrict the paths to
stay in the positive half of phase space as defined by the fermion
density matrix\cite{c103}. This would be a rigorous procedure if
we knew how to partition the space, but in practice one needs to
make an ansatz for the restriction.  A key unsolved problem for
these restricted path integrals is that the dynamics appears very
slow and non-ergodic at very low temperatures.

A key property for a Bose superfluid is the momentum distribution,
or its Fourier transform, the single particle off-diagonal density
matrix. This is obtained in path integrals by allowing one path to
be open, a linear ``polymer.'' The two open ends of this polymer
can become separated in a superfluid if that atom is a part of a
long permutation cycle.  Linear polymers have a very efficient way
to move through phase space, the reptation motion, (${\it i. e.}$
move like a slithering snake), developed for lattice and continuum
polymer simulations\cite{kron,web80}. The reptation algorithm is
obtained by cutting off one end of the polymer and growing that
part onto the other end while keeping the body of the snake and
the other polymers unchanged.  This is a very fast operation both
in computer time and in how quickly it refreshes the configuration
of the polymer. If one allows the length of the polymer to
fluctuate, growth at one end and shrinkage at the other, need not
be explicitly coupled. Real world polymers are polydisperse
anyway. The reptation algorithm is another example of a fruitful
insight in quantum algorithms coming from the polymer world.

One application of the reptation algorithm for quantum simulations
is to ground state path integral calculations\cite{bar99}. In the
ground state limit, closing of the paths becomes unimportant, and
instead one works with open polymers, closed on the ends with a
trial wavefunction. Since they are open, they can move by
reptation. This gets around the DMC problem of mixed estimators.

There has been considerable development of MCMC algorithms for
quantum lattice models. There are several crucial distinctions
between the lattice models and the continuum models, even for
bosonic systems. First of all, on the lattice, the action is
bounded, leading to other ways of approximating the action.
Secondly, there is a finite set of possible local moves, allowing
one to use heat bath methods. Suppose,we define the
``neighborhood'' of a state, as all states that can be reached by
a certain class of moves. By heat bath we mean that we directly
sample the equilibrium distribution in the neighborhood: $C_s
\pi(s)$. Finally, in lattice models the random walks are not
continuous trajectories. Some important principles such as
fixed-node and winding number estimators were discovered in the
continuum because they require continuous trajectories.

PIMC for a lattice model such as the Bose Hubbard model, is known
as ``world line Monte Carlo.'' For the reasons discussed above, it
has problems with convergence in the low temperature limit.
Progress\cite{evertz,kawa95} has made with loop and cluster moves,
as described  elsewhere in this volume by Troyer. These ideas have
given rise to the ``meron'' methods to solve the sign problem for
certain models\cite{chandra99}. In Prokofev's\cite{prok98} worm
algorithm, one starts with an open polymer and allows the two ends
to grow and shrink independently, as we described above with
reptation.  One gets correct equilibrium statistics (which require
closed loops) by taking averages over only those configurations
where the head and tail happen to land on the same sites.  These
new methods have allowed simulations of large lattices and
computation of critical properties of quantum phase transitions.

Lattice PIMC for fermion systems is referred to as determinantal
MC. There one performs a Stratonovitch-Hubbard transformation of
the interaction term, leading to the interaction of a Slater
determinant with a random field\cite{scalapino81}. Heat bath
algorithm and fermion update formulas are used in the
implementation of MCMC in this approach\cite{blank81}. Aside from
the half filled Hubbard model, one has a serious fermion sign
problem. Recent progress has been made in developing fixed-node
approaches\cite{zhang97,zhang03} for determinantal Monte Carlo.

\section{Metropolis when the energy function is random}

A significant generalization of the MCMC algorithm is the penalty
method. In most of the classical MC applications to date, it is
assumed that the energy function is computable in a finite number
of operations, and most applications before 1985 used an empirical
pair potential, with an occasional more complex functional form.
In 1985, Car-Parrinello\cite{car85}, showed that one can solve the
mean field density functional equations at each step in a
molecular dynamics simulation. However, to reach the accuracy
needed for many practical problems it will be necessary to go far
beyond mean field or semi-empirical approaches, greatly increasing
computer time.

An approach that we are following\cite{C168} is to calculate the
electronic Born-Oppenheimer energy  at zero temperature using a
DMC random walk. The ions  are moved at a non-zero temperature
with MCMC, possibly using multilevel sampling. However, this means
that the energy difference in the Metropolis acceptance formula
will not be known precisely but will have a statistical
fluctuation. For high accuracy, one will need to reduce these
fluctuations to much below $k_B T$ to get reliable results, but
how much lower does one need to go? The difference in computer
time in going from an error of $k_B T$ to $k_B T/10$ is a factor
of 100! Several years ago\cite{C139}, we raised the question of
whether it was possible to take into account these fluctuations in
the energy in the Metropolis algorithm. There have been a few
suggestions\cite{kennedy,krajci} about to handle noisy energy
evaluations in the past but without concern about large
fluctuations in the energy or of the efficiency of the approach.
Suppose that $\delta$ is an estimate of $\Delta E$ from a known
probability distribution, $P(\delta)$. That is $\int d\delta
P(\delta )\delta = \Delta E$. Let us require detailed balance on
the average:
\begin{equation}
  \int d\delta P(\delta)\left[ \pi(s) A(\delta)-\pi(s')A(-\delta)\right] = 0
\end{equation}
where for simplicity, we have assumed a symmetric sampling
function, $T(s \rightarrow s')=T(s' \rightarrow s)$. Assuming
reasonable conditions on the QMC evaluation of the energy,
$P(\delta)$ will approach a normal distribution with a variance
$\sigma^2$. Somewhat surprisingly, a nearly optimal solution to
the detailed balance equation for a normal distribution has been
discovered\cite{C139}. To satisfy detailed balance on average one
needs to accept a move with probability:
\begin{equation}
A(s \rightarrow s') = min [ 1 , \exp( -\beta\delta  -
\beta^2\sigma^2/2 )].
\end{equation}
One must add a {\it penalty} equal to $\beta^2\sigma^2/2$ to the
energy difference to compensate for the fluctuations. This is much
more efficient than simply beating down the error bars. The most
efficient simulation is one for which $\beta\sigma >1$: it is
better to take cheap moves, many of which are likely to be
rejected, rather than many fewer expensive moves with small values
of $\sigma^2$.

Clearly there are other situations aside from quantum simulations
where one might wish to evaluate the change in energy only
approximately. For example, one might imagine that a classical
energy can be split into large short-ranged terms, and slowly
varying long-range terms which are slow to evaluate. Those terms
can then be sampled. There are recent suggestions on how to use
this for protein folding and to deal with non-normally distributed
energy fluctuations\cite{ball}.

\section{Bayesian analysis}

In projector and path integral Monte Carlo, the ``dynamics'' is in
imaginary time.  By that is meant that we sample matrix elements
of $\exp[-\beta{\hat H}]$. An important problem is to extract a
maximum amount of information from the imaginary time correlations
of these simulations. For example, it is well known that real-time
linear response is related by a Laplace transform to the time
correlations in PIMC.  In this last application of MCMC, we
consider how it can be used in the data analysis. This application
of MCMC is completely different than the previous ones since the
random walk is not in the space of particle coordinates or
permutations, but in the space of the real time response.

In fact, we consider a related problem arising from the fermion
``sign''problem. The exact transient estimate
algorithm\cite{lee81} allows projection of the ground state for a
limited time, but at large time $\beta$, the estimates get
increasingly noisy.  The simple solution is to take the largest
time projection as the best estimate of the fermion energy.
However, it is clear that there is more statistical information if
the earlier values of $\beta$ are also used\cite{C070}.  In
transient estimate MC, we determine:
\begin{equation}
 h(t) = \langle \Psi_T | exp[-\beta{\hat H}] \Psi_T \rangle
 \end{equation}
as well as the time derivatives of $h(t)$ for a range of times $0
\leq t \leq \beta$. As $\beta$ gets large, the signal to noise
ratio for the energy goes exponentially to zero.  But $h(t)$ is
related to the spectrum of ${\hat H}$ by:
 \begin{equation}
 h(t) = \int_{-\infty}^{+\infty} dE c(E) e^{-t E}
 \label{LPT}\end{equation}
where the spectral density is
 \begin{equation}
 c(E) = \sum_i \delta(E-E_i)| \langle \Psi_T | \phi_i \rangle |^2
 \end{equation}
and $(E_i,\phi_i)$ are the exact energies and wavefunctions for
state $i$. Some analytic information is known about the spectral
density, namely, that it is positive, is identically zero for $E<
E_0$ and decays at large $E$ as $E^{-k}$. Since the ``Laplace''
transform in Eq.\ref{LPT} is a smoothing operation, the inverse
transform needed to find $c(E)$ from $h_{MC}(t)$ is
ill-conditioned. Because the evaluation of $h_{MC}(t)$ is
stochastic, there is a distribution of $c(E)$'s, all of which are
consistent with the DMC-determined data. Taking the Bayesian point
of view, there is a prior distribution of $Pr_m [ c] $ which is
conventionally chosen to be an entropic function. Then the
posterior distribution of $c(E)$ is given by the Bayesian formula:
 \begin{equation}
    Pr(c | h_{MC} ) \propto  Pr_L(h_{MC} | c ) Pr_m [ c ]
 \end{equation}
where $Pr_l (h_{MC} |c)$, the likelihood function, is the
probability of obtaining the Monte Carlo data assuming a given
spectral density, $c(E)$.  By the central limit theorem, the
likelihood function is a multivariate Gaussian, whose parameters
we can estimate within DMC.

The maximum entropy method, MAXENT\cite{jarrell} consists of
finding the most probable value of $c(E)$ and estimating errors by
expanding around the maximum. However, MCMC can be put to good use
in sampling the distribution of $c(E)$, particularly for cases
where the overall distribution $Pr(c | h_{MC} )$ is non-gaussian.
To do this one represents $c(E)$ on a finite grid, considers moves
that change the values of $c(E)$ and accepts or rejects such moves
based on how the move changes the value of $Pr(c | h_{MC} )$. The
structure of this distribution function is different than those
that arise in simulations of particle or lattice systems. For
example, the moving distance for $c(E)$ should depend on $E$.
Though this analysis takes longer than MAXENT to estimate the
spectrum, it typically takes much less time than the original DMC
calculation that generated $h_{MC} (t)$. By looking at the output
of the MCMC sampling of $c(E)$ one can get a quantitatively
precise estimate of which spectral reconstructions are likely.
More physical prior functions or analytic insight are easy to put
into the distribution function. The density-density response
function ($S_k (\omega)$) of liquid $^4$He\cite{C111} has been
calculated using this Metropolis procedure. These statistical
estimation methods are increasingly used in computational
statistics\cite{MCMC}, under the acronym MCMC and are more fully
described elsewhere in this volume.


\begin{theacknowledgments}
This research was funded by NSF DMR01-04399 and the Dept. of
Physics at the University of Illinois.  Much of the early work
described here was done in collaboration with G. V. Chester and M.
H. Kalos. M. H. Kalos is also acknowledged for comments on an
early version of this manuscript.
\end{theacknowledgments}

\bibliographystyle{aipproc}   

\bibliography{dmc,BIBFILE}

\end{document}